\documentclass[prc,showpacs,floatfix]{revtex4}
\usepackage{bm}
\usepackage{dcolumn}
\usepackage{graphicx}
\usepackage{amsmath}
\usepackage{bbold}
\newcommand{\beq}{\begin{equation}}
\newcommand{\eeq}{\end{equation}}
\newcommand{\beqa}{\begin{eqnarray}}
\newcommand{\eeqa}{\end{eqnarray}}
\begin{document}
\title{
\hfill{\small {\bf MKPH-T-05-05}}\\
{\bf Incoherent pion photoproduction on the deuteron with
polarization observables I: Formal expressions}} 
\author{H. Arenh\"ovel and A. Fix }
\affiliation{
Institut f\"ur Kernphysik,
Johannes Gutenberg-Universit\"at Mainz, D-55099 Mainz, Germany}
\date{\today}
\begin{abstract}
Formal expressions are developed for the general five-fold differential 
cross section of incoherent $\pi$-photoproduction on the deuteron
including beam and target polarization. The polarization observables 
of the cross section are described by various beam, target and
beam-target asymmetries for polarized photons and/or polarized
deuterons. They are given as bilinear
hermitean forms in the reaction matrix elements divided by the
unpolarized cross section. In addition, the corresponding observables
for the semi-exclusive reaction $\vec d(\vec \gamma,\pi)NN$ are also
given. 
\end{abstract}

\maketitle

\section{Introduction}
Photoproduction of pions on light nuclei is an important topic in
medium energy nuclear physics. It is motivated by different and 
complementary aspects. On the one hand one wants to study the
elementary reaction on the neutron to which otherwise one has no
access. On the other hand one is interested in the influence of a 
nuclear environment on the elementary production amplitude, and 
last but not least, one hopes to obtain information on nuclear 
structure.

Besides the study of unpolarized total and differential cross
sections, polarization observables provide very often further insight
into details of the underlying reaction mechanisms and possible
structure effects. In this case, such observables will serve as
additional critical tests or check points for theoretical models. The
considerable progress in experimental techniques for studying
polarization phenomena has brought into focus also the question, what
role polarization effects play in pion photoproduction on nuclei. Of
particular interest is photoproduction of pions on the deuteron in
view of its simple structure. Indeed, it has been studied quite
extensively over the past 50 years (see~\cite{DaA03a} and references
therein). While in earlier work mainly total and semi-exclusive
differential cross sections of incoherent pion production have been
studied, polarization observables were considered more recently, both
in experiment~\cite{LEGS,A2} as well as in theory. For example, the
spin asymmetry of the total cross section with respect to circular
photon polarization, which determines the Gerasimov-Drell-Hearn sum 
rule, was investigated theoretically in~\cite{LeP96,DaA03b,ArF04}
and target asymmetries were considered in~\cite{LoS00}. 

Subsequently, various polarization asymmetries of the semi-exclusive
differential cross section $\vec d(\vec\gamma,\pi)NN$ 
were studied theoretically in a series of
papers~\cite{Dar04a,Dar05a,Dar05b,Dar05c,DaS05}. Unfortunately, many
of the results presented there are based on incorrect expressions for
polarization observables, because the formal expressions for them were
taken in analogy from the corresponding expressions of deuteron
photodisintegration~\cite{Are88}. This is in principle possible, since
the spin degrees of freedom are the same in both reactions, provided
one takes care to check where certain symmetry properties of the
reaction amplitude have been used in the derivation of the
polarization observables in photodisintegration, because they are not
identical in both reactions. This caveat refers in particular to those
observables which are related to linearly polarized photons. It
appears that this fact was not taken into account so that the 
results in~\cite{Dar05a,Dar05b} for them cannot be trusted. But also
the results for circularly polarized photons are incorrect, namely the
claim in~\cite{Dar04a}, that all of them vanish identically, is
wrong. Moreover, this statement is in contradiction to~\cite{Dar05b},
where a non-vanishing differential spin asymmetry for circularly
polarized photons is reported, because this asymmetry is proportional
to the beam-target asymmetry $T^c_{10}$ for circularly polarized photons 
and a vector polarized deuteron, which means that the latter does
not vanish. Thus, it is obvious that the importance of polarization
effects requires a more careful and thorough treatment as done
in~\cite{Dar04a,Dar05a,Dar05b,Dar05c,DaS05}. 

With the present work we want to provide a solid basis for the formal
expressions of the various polarization observables which determine
the differential cross section for incoherent pion production on the
deuteron with polarized photons and/or polarized deuterons by deriving
the general form of the differential cross section including all
possible polarization asymmetries. It complements the work of Blaazer
et al.~\cite{BlB94}, who have formally derived all possible
polarization observables for coherent pion photoproduction on the
deuteron. 

\section{Kinematics}

\begin{figure}[htb]
\includegraphics[scale=.6]{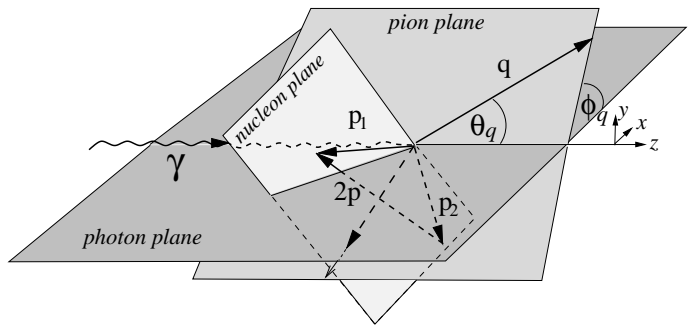}
\caption{Kinematics of pion photoproduction on the deuteron in the 
laboratory system.}
\label{fig_kinematics}
\end{figure}

As a starting point, we will first consider the kinematics of 
the photoproduction reaction 
\beq
\gamma(k,\vec{\varepsilon}_\mu)+d(p_d)\!\rightarrow\!
\pi(q)+N_1(p_1)+N_2(p_{2})\,,
\eeq
where we have defined the notation of the four-momenta of the participating
particles. The circular polarization vector of the photon is denoted by
$\vec{\varepsilon}_\mu$ with $\mu=\pm 1$. The following formal
developments will not depend on the reference frame, laboratory or
center-of-momentum (c.m.) frame. However, in view of our explicit
application~\cite{FiA05} in which the reaction is evaluated in the 
laboratory frame, we will refer sometimes to this frame for
definiteness. We choose as independent variables for the 
description of the final state
the outgoing pion momentum $\vec q=(q,\theta_q,\phi_q)$ and the
spherical angles $\Omega_p=(\theta_p,\phi_p)$ of the relative momentum
$\vec p=(\vec p_1-\vec p_2)/2=(p,\Omega_p)$ of the two outgoing
nucleons. Together with the incoming photon energy $\omega=k_0$, 
the momenta of the outgoing nucleons are fixed, i.e.\
$\vec p_{1/2}=(\vec k +\vec p_d -{\vec q})/{2} \pm \vec p$.
The coordinate system is chosen to have a right-handed
orientation with $z$-axis along the photon momentum $\vec{k}$. We
distinguish in general three planes: (i) the photon plane spanned by
the photon momentum and the direction of maximal linear photon
polarization, which defines the direction of the $x$-axis, (ii) the
pion plane, spanned by the photon and pion momenta, which intersects the
photon plane along the $z$-axis with an angle $\phi_q$, and (iii) the
nucleon plane spanned by the momenta of the two outgoing nucleons 
intersecting the pion plane along the total momentum of the two
nucleons. This is illustrated in Fig.~\ref{fig_kinematics} for the
laboratory frame. In case that the linear photon polarization vanishes, one can
choose $\phi_q=0$ and then photon and pion planes coincide.

\section{The $T$-matrix}

All observables are determined by the $T$-matrix elements of the
electromagnetic pion production current $\vec J_{\gamma\pi}$ between the
initial deuteron and the final $\pi NN$ states. In a general frame, it
is given by
\beq
T_{s m_s, \mu m_d}= -^{(-)}\langle \vec p_1\,\vec p_2\,s
m_s,\,\vec q\,| \vec\varepsilon_\mu\cdot\vec J_{\gamma\pi}(0)|
\vec p_d\,1 m_d\rangle\,,
\eeq
where $s$ and $m_s$ denote the total spin and its projection on the
relative momentum $\vec p$ of the outgoing two nucleons, and $m_d$ 
correspondingly the deuteron spin projection on the $z$-axis as 
quantization axis. Furthermore, transverse gauge has been chosen. 
The knowledge of the specific form of $\vec J_{\gamma\pi}$ is not 
needed for the following formal considerations. 

The general form of the $T$-matrix after separation of the overall
c.m.-motion is given by 
\begin{eqnarray}
T_{s m_s \mu m_d}(q,\,\Omega_q,\,\Omega_p)&=& -
^{(-)}\langle \vec p\, s m_s,\,\vec q\,|J_{\gamma\pi,\,\mu}(\vec k\,)|1
m_d\rangle\nonumber\\ 
&=& \sqrt{2\pi}\sum_{L} i^L\hat L \,
^{(-)}\langle \vec p\, s m_s,\,\vec q\,|{\cal O}^{\mu L}_\mu|1 m_d\rangle
\end{eqnarray}
with $\mu=\pm 1$ and transverse electric and magnetic multipoles
\begin{eqnarray}
{\cal O}^{\mu L}_M&=& E_M^L +\mu M_M^L\,.
\end{eqnarray}
Furthermore, we use throughout the notation $\hat L=\sqrt{2L+1}$. 
It is convenient to introduce a partial wave decomposition of the
final states by
\begin{eqnarray}
^{(-)}\langle \vec p\, s m_s|&=&\frac{1}{\sqrt{4\pi}}
\sum_{l_p j_p m_p}\hat l_p(l_p 0 s m_s|j_p m_s)\,
D^{j_p}_{m_s,m_p}(\phi_p,-\theta_p,-\phi_p)
^{(-)}\langle p (l_p s)j_p m_p|\,,\\
^{(-)}\langle \vec q\,|&=&\frac{1}{\sqrt{4\pi}}
\sum_{l_q m_q}\hat l_q\,D^{l_q}_{0,m_q}(\phi_q,-\theta_q,-\phi_q)
^{(-)}\langle q l_q m_q|\,,
\end{eqnarray}
where $m_p$ and $m_q$ like $m_d$ refer to the photon momentum $\vec k$ 
as quantization axis.
Here, the rotation matrices $D^j_{m'm}$ are taken in the convention of
Rose~\cite{Ros57}. Using the multipole decomposition and applying the
Wigner-Eckart theorem yields
\begin{eqnarray}
^{(-)}\langle p (l_p s)j_p m_p, q\,l_q m_q|{\cal O}^{\mu L}_M|1 m_d\rangle
&=&\sum_{J M_J}(-)^{j_p-l_q+J}\,\hat{J}
\left(\begin{array}{ccc} 
j_p & l_q & J \cr m_p & m_q & -M_J\cr
\end{array}\right)
\left(\begin{array}{ccc} 
J & L & 1 \cr -M_J & M & m_d\cr
\end{array}\right)\nonumber\\&&\hspace*{2.3cm}
\langle p q ((l_ps)j_p l_q)J||{\cal O}^{\mu L}||1\rangle\,,
\end{eqnarray}
with the selection rule $m_p+m_q=M_J=M+m_d$. Rewriting the angular
dependence 
\begin{eqnarray}
D^{j_p}_{m_s,m_p}(\phi_p,-\theta_p,-\phi_p)\,
D^{l_q}_{0,m_q}(0,-\theta_q,-\phi_q)&=&
d^{j_p}_{m_s,m_p}(-\theta_p)\,d^{\,l_q}_{0,m_q}(-\theta_q)\,
e^{i((m_p-m_s)\phi_p+m_q\phi_q)}\,,
\end{eqnarray}
and rearranging, using the foregoing selection rule for $M=\mu$,
\begin{eqnarray}
(m_p-m_s)\phi_p+m_q\phi_q&=&(m_p-m_s)\phi_{pq}+(\mu+m_d-m_s)\phi_q
\end{eqnarray}
with $\phi_{pq}=\phi_{p}-\phi_{q}$, one finds that the $T$-matrix can
be written as 
\begin{eqnarray}\label{small_t}
T_{s m_s \mu m_d}(\Omega_p,\Omega_q)&=& e^{i(\mu+m_d-m_s)\phi_q}
t_{s m_s \mu m_d}(\theta_p,\,\theta_q,\,\phi_{pq})\,,
\end{eqnarray}
where the small $t$-matrix depends only on $\theta_p$, $\theta_q$, and
the relative azimuthal angle $\phi_{pq}$. Explicitly one has
\begin{eqnarray}
t_{s m_s \mu m_d}(\theta_p,\,\theta_q,\,\phi_{pq})&=&
\frac{1}{2\,\sqrt{2\pi}}
\sum_{L l_p j_p m_p l_q m_q J J M_J}i^L\,\hat L\,\hat J\,\hat l_q
\,\hat l_p\,\hat j_p\,(-)^{J+l_p+j_p-s+m_s-l_q}\nonumber\\
&&\left(\begin{array}{ccc} 
l_p & s & j_p \cr 0 & m_s & -m_s\cr
\end{array}\right)
\left(\begin{array}{ccc} 
j_p & l_q & J \cr m_p & m_q & -M_J\cr
\end{array}\right)
\left(\begin{array}{ccc} 
J & L & 1 \cr -M_J & \mu & m_d\cr
\end{array}\right)\nonumber\\
&&\langle p q ((l_ps)j_p l_q)J||{\cal O}^{\mu L}||1\rangle
d^{j_p}_{m_s,m_p}(-\theta_p)\,d^{l_q}_{0,m_q}(-\theta_q)\,
e^{i(m_p-m_s)\phi_{pq}}\,.\label{smallt}
\end{eqnarray}

Using this explicit form for the small $t$-matrix, it is quite
straightforward to show that, if parity is conserved, the following
symmetry relation holds for the inverted spin projections 
\begin{eqnarray}\label{symmetry}
t_{s -m_s -\mu -m_d}(\theta_p,\,\theta_q,\,\phi_{pq})&=&
(-)^{s+m_s+\mu+m_d}
t_{s m_s \mu m_d}(\theta_p,\,\theta_q,\,-\phi_{pq})\,.
\end{eqnarray}
In the derivation of this relation one has made use of the parity 
selection rules for the multipole transitions to a final partial 
wave $|p q ((l_ps)j_p l_q)J\rangle$ with parity 
$\pi_{J(l_p,l_q)}=(-)^{l_p+l_q+1}$
\begin{eqnarray}
\left\{\begin{array}{lll} 
E^L & \pi_d \pi_{J(l_p,l_q)}\,(-)^L=1 &\rightarrow (-)^{l_p+l_q+L}=-1\cr
M^L & \pi_d \pi_{J(l_p,l_q)}\,(-)^L=-1 &\rightarrow (-)^{l_p+l_q+L}=1\cr
\end{array}\right\}\,.
\end{eqnarray}
Therefore, invariance under a parity transformation results in the
following property of the reduced matrix element 
\begin{eqnarray}
(-)^{l_p+l_q+L}\langle p q ((l_ps)j_p l_q)J||{\cal O}^{-\mu L}||1\rangle&=&
-\langle p q ((l_ps)j_p l_q)J||{\cal O}^{\mu L}||1\rangle\,.
\end{eqnarray}
The symmetry property (\ref{symmetry}) 
leads to a corresponding relation for the $T$-matrix
\begin{eqnarray}
T_{s -m_s -\mu -m_d}(\theta_p,\phi_p,\theta_q,\phi_q)&=&
(-)^{s+m_s+\mu+m_d}
T_{s m_s \mu m_d}(\theta_p,-\phi_p,\theta_q,-\phi_q)\,.
\end{eqnarray}
For an uncoupled spin representation, one finds accordingly, using the 
transformation
\begin{eqnarray}
T_{m_1 m_2 \mu m_d}(\theta_p,\phi_p,\theta_q,\phi_q)&=&
\sum_{s m_s}(\frac{1}{2} m_1 \frac{1}{2} m_2|s m_s)\,
T_{s m_s \mu m_d}(\theta_p,\phi_p,\theta_q,\phi_q)\,,
\end{eqnarray}
where $m_j$ denotes the spin projection of the ``jth'' nucleon on 
the quantization axis, as symmetry relation
\begin{eqnarray}
T_{-m_1 -m_2 -\mu -m_d}(\theta_p,\phi_p,\theta_q,\phi_q)&=&
(-)^{1+m_1+m_2+\mu+m_d}
T_{m_1 m_2 \mu m_d}(\theta_p,-\phi_p,\theta_q,-\phi_q)\,.
\end{eqnarray}

The small $t$-matrix elements are the basic quantities which determine
differential cross section and asymmetries. The latter are given as 
ratios of bilinear hermitean forms in terms of the $t$-matrix elements (see
(\ref{vim}) and (\ref{wim}) below). 

\section{The differential cross section including polarization asymmetries}
\label{diff_cross}

The usual starting point is the general expression for the
differential cross section 
\beqa
\frac{d^5\sigma}{dq d\Omega_q d\Omega_p}&=&c(\omega,q,\Omega_q,\Omega_p)\,
tr(T^\dagger T\rho_i)\,,\label{diffcrossx}
\eeqa
where $T$ denotes the reaction matrix, $\rho_i$ the density 
matrix for the spin degrees of the initial system. 
The trace refers to all initial and final state spin degrees of freedom 
comprising incoming photon, target deuteron, and final nucleons. Furthermore, 
$c(\omega,q,\Omega_q,\Omega_p)$ denotes a kinematic factor which
comprises the final state phase space and the incoming flux. In an
arbitrary frame one has
\beq
c(\omega,q,\Omega_q,\Omega_p)=\frac{1}{(2\pi)^5}\,\frac{E_d}{E_d+p_d}\,
\frac{m_N^2}{4\omega\omega_\pi}\,\frac{p^*\,q^2}{W_{NN}}\,
\eeq
with 
\beq
p^*=\frac{1}{2}\,\sqrt{W_{NN}^2-4\,m_N^2}\,,
\eeq
as the relative momentum of the final two nucleons in their c.m.\ sytem, and 
\beq
\omega=k_0\,,\quad E_d=\sqrt{p_d^2+m_d^2}\,,\quad 
\omega_\pi=\sqrt{q^2+m_\pi^2}\,,
\quad W_{NN}^2=(\omega +E_d- \omega_\pi)^2-(\vec k+\vec p_d-\vec q\,)^2\,.
\eeq 
The density matrix $\rho_i$ in (\ref{diffcrossx}) is a
direct product of the density matrices $\rho^\gamma$ of the 
photon and $\rho^d$ of the deuteron 
\begin{equation}
\rho_i=\rho^\gamma\otimes\rho^d \,.
\end{equation}
The photon density matrix has the form
\beq
\rho^\gamma_{\mu \mu'}=\frac{1}{2}(\delta_{\mu \mu'}+
\vec P^\gamma\cdot\vec \sigma_{\mu \mu'})
\eeq
with respect to circular polarization $\mu=\pm 1$. Here, $|\vec
P^\gamma|$ describes the total degree of polarization,
$P^\gamma_z=P^\gamma_c$ the degree of circular polarization, and
$P^\gamma_l=\sqrt{(P^\gamma_x)^2+(P^\gamma_y)^2}$ the degree of linear
polarization. By a proper rotation around the photon momentum, one can
choose the $x$-axis in the direction of maximum linear polarization,
i.e., $P^\gamma_x=-P^\gamma_l$ and $P^\gamma_y=0$. Then one has explicitly
\beq
\rho^\gamma_{\mu \mu'}=(1+\mu\,P^\gamma_c)\,\delta_{\mu \mu'}-P^\gamma_l\,
\delta_{\mu, -\mu'}\,e^{2i\mu\phi_q}\,.
\eeq

Furthermore, the deuteron density matrix $\rho^d$ can be expressed in terms of 
irreducible spin operators $\tau^{[I]}$ with respect to the 
deuteron spin space 
\begin{equation}
\rho_{m_d\, {m_d}'}^d=\frac{1}{3}
\sum_{I\,M}(-)^M\hat{I}\,\langle 1m_d|\tau^{[I]}_M |1m_d' \rangle
P^d_{I-M}\,,\label{rhod}
\end{equation}
where $P^d_{00}=1$, and $P^d_{1M}$ and $P^d_{2M}$ describe vector 
and tensor polarization components of the deuteron, respectively. 
The spin operators are defined by their reduced matrix elements
\begin{equation}
\langle 1||\tau^{[I]}||1 \rangle = \sqrt{3}\,\hat I
\quad\mbox{for}\quad I=0,1,2\,.
\end{equation}
From now on we will assume that the deuteron density matrix is diagonal 
with respect to an orientation axis $\vec d$ having spherical angles 
$(\theta_d,\phi_d)$ with respect to the coordinate system associated with 
the photon plane in the lab frame. Then one has with respect to
$\vec d$ as quantization axis 
\begin{equation}
\rho_{m\,m'}^d=p_m\,\delta_{m\,m'}\,,
\end{equation}
where $p_m$ denotes the probability for finding a deuteron spin projection $m$ 
on the orientation axis. With respect to this axis one finds from (\ref{rhod}) 
$P^d_{I\,M}(\vec d\,)=P^d_I\,\delta_{M,0}$, where the orientation 
parameters $P_I^d$ are related to the probabilities $\{p_m\}$ by
\beqa
P_I^d&=&\sqrt{3}\,\hat{I}\sum_{m}(-)^{1-m}
\left( 
\begin{matrix}
1&1&I \cr m &-m & 0 \cr
\end{matrix} \right)p_m\nonumber\\
&=& \delta_{I 0} + \sqrt{\frac{3}{2}}(p_1-p_{-1})\,\delta_{I 1} 
+\frac{1}{\sqrt{2}}\,(1-3\,p_0)\,\delta_{I 2}\,.
\eeqa
The polarization components in the chosen lab frame are obtained 
from the $P^d_I$ by a rotation, transforming the quantization axis along 
the orientation axis into the direction of the photon momentum, i.e.\
\beq
P^d_{IM}(\vec z\,)=P^d_Ie^{iM\phi_d}d^I_{M0}(\theta_d)\,,
\eeq
where $d^j_{m m'}$ denotes a small rotation matrix~\cite{Ros57}.
Thus the deuteron density matrix becomes finally
\begin{equation}
\rho_{m_d\, {m_d}'}^d=\frac{1}{\sqrt{3}}(-)^{1-m_d}
\sum_{I\,M}\hat{I}
\left( 
\begin{matrix}
1&1&I \cr m_d'&-m_d&M \cr
\end{matrix} \right) P_I^d
e^{-iM\phi_d}d^I_{M0}(\theta_d)\,. \label{rhoda}
\end{equation}
This means, 
the deuteron target is characterized by four parameters, namely the 
vector and tensor polarization parameters $P_1^d$ and $P_2^d$, respectively,
and by the orientation angles $\theta_d$ and $\phi_d$.  If one chooses
the c.m.\ frame as reference frame, one should note that the 
deuteron density matrix undergoes no change in the 
transformation from the lab to the c.m. system, since the boost to the c.m.\ 
system is collinear with the deuteron quantization axis~\cite{Rob74}.

The evaluation of the general expression of the differential
cross section in (\ref{diffcrossx}) can be done analogously to deuteron
photodisintegration as described in detail in~\cite{Are88}. In fact,
one can follow the same steps except for the use of the symmetry
relation of Eq.~(2) in~\cite{Are88} which is different in case of pion
production (see (\ref{symmetry})) because of the additional pion
degree of freedom in the final state, in particular its pseudovector 
character. In terms of the small
$t$-matrices as defined in (\ref{small_t}), one finds, inserting the
density matrices of photon and deuteron for the general five-fold 
differential cross section,
\beqa
\frac{d^5\sigma}{ dq d\Omega_q
d\Omega_p}&=&\frac{1}{2}
\sum_{\mu'\mu IM} P^d_I \,e^{iM(\phi_q-\phi_d)}d^I_{M0}(\theta_d)
u_{IM}^{\mu'\mu}\,\Big[(1+\mu\,P^\gamma_c)\delta_{\mu \mu'}-P^\gamma_l\,
\delta_{\mu, -\mu'}e^{2i\mu\phi_q}\Big]\,,
\eeqa
where we have introduced the quantities 
\beqa
u_{IM}^{\mu'\mu}(q,\, \theta_q,\, \theta_p,\, \phi_{pq})&=& 
c(\omega,q,\Omega_q,\Omega_p)\,
\frac{\hat I}{\sqrt{3}}\,\sum_{m_d m_d'}(-)^{1-m_d}
\left( 
\begin{matrix}
1&1&I \cr m_d'&-m_d&M \cr
\end{matrix} \right)\nonumber\\
&& 
\sum_{s m_s}t^*_{s m_s \mu' m_d'}(q,\, \theta_q,\, \theta_p,\, \phi_{pq})
\,t_{s m_s \mu m_d}(q,\, \theta_q,\, \theta_p,\,
\phi_{pq})\,.\label{uim}
\eeqa
It is straighforward to prove that they behave under complex
conjugation as 
\beq\label{complc}
u_{IM}^{\mu'\mu}(q,\, \theta_q,\, \theta_p,\, \phi_{pq})^*=
(-)^M\,u_{I-M}^{\mu\mu'}(q,\, \theta_q,\, \theta_p,\, \phi_{pq})\,.
\eeq
Furthermore, with the help of the symmetry in (\ref{symmetry}) one finds
\beq
u_{IM}^{-\mu'-\mu}(q,\, \theta_q,\, \theta_p,\, \phi_{pq})=
(-)^{I+M+\mu'+\mu}\,u_{I-M}^{\mu'\mu}(q,\, \theta_q,\, \theta_p,\,
-\phi_{pq})\,, 
\eeq
which yields in combination with (\ref{complc})
\beq\label{complca}
u_{IM}^{-\mu'-\mu}(q,\, \theta_q,\, \theta_p,\, \phi_{pq})=
(-)^{I+\mu'+\mu}\,u_{IM}^{\mu\mu'}(q,\, \theta_q,\, \theta_p,\,
-\phi_{pq})^*\,.
\eeq
This relation is quite useful for a further simplification of the
semi-exclusive differential cross section later on.

Separating the polarization parameters of photon ($P^\gamma_l$ and
$P^\gamma_c$) and deuteron ($P^d_I$), it is then straightforward to
show that the differential cross section can be brought into the form 
\beqa
\frac{d^5\sigma}{ dq d\Omega_q d\Omega_p }&=&
\frac{1}{2}\sum_{I}P^d_I \,
\sum_{M=-I}^I \,e^{iM\phi_{qd}}\,
d^I_{M0}(\theta_d)\,\Big[v_{IM}^1 +v_{IM}^{-1}\nonumber\\&& 
+ P^\gamma_c\,(v_{IM}^1-v_{IM}^{-1})
+P^\gamma_l\,(w_{IM}^1\,e^{-2i\phi_{q}}+w_{IM}^{-1}\,e^{2i\phi_{q}})
\Big]\,,\label{diffcrossd}
\eeqa
with $\phi_{qd}=\phi_{q}-\phi_{d}$, where we have introduced for
convenience the quantities 
\beqa
v_{IM}^\mu(q,\, \theta_q,\, \theta_p,\, \phi_{pq})&=& 
u_{IM}^{\mu\mu}(q,\, \theta_q,\, \theta_p,\, \phi_{pq})\,,\label{vim}\\
w_{IM}^\mu(q,\, \theta_q,\, \theta_p,\, \phi_{pq})&=& 
-u_{IM}^{\mu \,-\mu}(q,\, \theta_q,\, \theta_p,\, \phi_{pq})\,.\label{wim}
\eeqa
According to (\ref{complc}) and (\ref{complca}), they have the
following properties under complex conjugation 
\beqa
v/w_{IM}^\mu(q,\, \theta_q,\, \theta_p,\, \phi_{pq})^*&=&(-)^M
v/w_{I-M}^\mu(q,\, \theta_q,\, \theta_p,\, \phi_{pq})\,,\label{vw}\\
v_{IM}^{\mu}(q,\, \theta_q,\, \theta_p,\, \phi_{pq})^*&=&(-)^I
v_{IM}^{-\mu}(q,\, \theta_q,\, \theta_p,\, -\phi_{pq}) \,,\label{vminus}\\
w_{IM}^\mu(q,\, \theta_q,\, \theta_p,\, \phi_{pq})^*&=&(-)^I w_{IM}^\mu(q,\,
\theta_q,\, \theta_p,\, -\phi_{pq})\,.\label{wima} 
\eeqa
From Eq.~(\ref{vw}) follows that $v_{I0}^{\mu}$ and $w_{I0}^\mu$ are real.
The sum over $M$ in (\ref{diffcrossd}) can be rearranged with the help
of the relation (\ref{vw}) and $d^I_{-M0}(\theta_d)=(-)^Md^I_{M0}(\theta_d)$
\beqa
\sum_{M=-I}^I \,e^{iM\phi_{qd}}\,
d^I_{M0}(\theta_d)\,(v_{IM}^1 \pm v_{IM}^{-1})&=&
\sum_{M=0}^I \frac{d^I_{M0}(\theta_d)}{1+\delta_{M0}}\,\Big(e^{iM\phi_{qd}}\,
(v_{IM}^1 \pm v_{IM}^{-1})+
e^{-iM\phi_{qd}}\,(-)^M
\,(v_{I-M}^1 \pm v_{I-M}^{-1})\Big)\nonumber\\
&=&\sum_{M=0}^I \frac{d^I_{M0}(\theta_d)}{1+\delta_{M0}}\,
\Big(e^{iM\phi_{qd}}\,(v_{IM}^1 \pm v_{IM}^{-1})+\mbox{c.c.}\Big)\,,
\eeqa
and furthermore with $\psi_M=M\phi_{qd}-2\,\phi_{q}$ 
\beqa
\sum_{M=-I}^I \,e^{iM\phi_{qd}}\,
d^I_{M0}(\theta_d)\,(w_{IM}^1\,e^{-2i\phi_q}
+ w_{IM}^{-1}\,e^{2i\phi_q})&=&
\sum_{M=-I}^I d^I_{M0}(\theta_d)\,\Big(e^{i\psi_M}\,
w_{IM}^1 +e^{-i\psi_M}\,(-)^M\,w_{I-M}^{-1}\Big)\nonumber\\
&=&\sum_{M=-I}^I d^I_{M0}(\theta_d)\,
\Big(e^{i\psi_M}\,w_{IM}^1+\mbox{c.c.}\Big)\,. 
\eeqa
This then yields for the differential cross section 
\beqa
\frac{d^5\sigma}{ dq d\Omega_q d\Omega_p}&=&
\sum_{I}P^d_I \,
\Big\{ \sum_{M=0}^I \frac{1}{1+\delta_{M0}}\,d^I_{M0}(\theta_d)
\Re e \,[e^{iM\phi_{qd}}\,(v_{IM}^+ +P^\gamma_c\,v_{IM}^-)]
\nonumber\\&& 
+\,P^\gamma_l\,\sum_{M=-I}^I d^I_{M0}(\theta_d)\,
\Re e \,[e^{i\psi_M}w_{IM}^1]
\Big\}\,,\label{diffcrossa}
\eeqa
where we have defined
\beq
v_{IM}^\pm=v_{IM}^1\pm v_{IM}^{-1}\,.
\eeq
Now, introducing various beam, target and beam-target asymmetries by
\beqa
\tau^{0/c}_{IM}(q,\, \theta_q,\, \theta_p,\, \phi_{pq})&=&
\frac{1}{1+\delta_{M0}}\,
\Re e \,v_{IM}^\pm(q,\, \theta_q,\, \theta_p,\, \phi_{pq})\,,\quad
M\ge 0\,,\label{tau0c}\\
\sigma^{0/c}_{IM}(q,\, \theta_q,\, \theta_p,\, \phi_{pq})&=&
-\Im m \,v_{IM}^\pm(q,\, \theta_q,\, \theta_p,\, \phi_{pq})\,,\quad
M> 0\,,\\
\tau^{l}_{IM}(q,\, \theta_q,\, \theta_p,\, \phi_{pq})&=&
\Re e \,w_{IM}^1(q,\, \theta_q,\, \theta_p,\, \phi_{pq})\,,\\
\sigma^{l}_{IM}(q,\, \theta_q,\, \theta_p,\, \phi_{pq})&=&
-\Im m \,w_{IM}^1(q,\, \theta_q,\, \theta_p,\, \phi_{pq})\,,\quad
M\neq 0\,,\label{sigmal}
\eeqa
where we took into account that $v_{I0}^{\mu}$ and $w_{I0}^\mu$ are
real, one obtains as final expression for the general five-fold
differential cross section with beam and target polarization
\beqa
\frac{d^5\sigma}{ dq d\Omega_q d\Omega_p}&=&
\sum_{I} P^d_I \,
\Big\{ \sum_{M=0}^I d^I_{M0}(\theta_d)
\Big[\tau^{0}_{IM}\,\cos{(M\phi_{qd})}
+\sigma^{0}_{IM}\,\sin{(M\phi_{qd})}\nonumber\\&& 
+P^\gamma_c\,(\tau^{c}_{IM}\,\cos{(M\phi_{qd})}
+\sigma^{c}_{IM}\,\sin{(M\phi_{qd})})\Big]
\nonumber\\&& 
+\,P^\gamma_l\,\sum_{M=-I}^I d^I_{M0}(\theta_d)\,
\Big[\tau^{l}_{IM}\,\cos{\psi_M}+\sigma^{l}_{IM}\,\sin{\psi_M}\Big]
\Big\}\,.\label{diffcross}
\eeqa
This constitutes our central result. 

We will now turn to the semi-exclusive reaction $\vec
d(\vec\gamma,\pi)NN$ where only the produced pion is detected, which
means integration of the five-fold differential cross section
$d^5\sigma/dq d\Omega_q d\Omega_p$ over $\Omega_p$. The resulting
cross section will then be governed by the integrated asymmetries
$\int d\,\Omega_{p}\,\tau_{IM}^\alpha$ and $\int
d\,\Omega_{p}\,\sigma_{IM}^\alpha$ ($\alpha\in\{0,c,l\}$), of which
quite a few will vanish, either $\int d\,\Omega_{p}\,\tau_{IM}^\alpha$
or $\int d\,\Omega_{p}\,\sigma_{IM}^\alpha$. To show this, we first
introduce the quantities 
\beqa
W_{IM}(q,\, \theta_q)&=&
\int d\,\Omega_{p}\,w_{IM}^{1}(q,\, \theta_q,\, \theta_p,\, \phi_{pq})
\nonumber\\&=& 
-\frac{\hat I}{\sqrt{3}}\,\int d\,\Omega_{p}\,c(\omega,q,\Omega_q,\Omega_p)\,
\sum_{m_d m_d'}(-)^{1-m_d}
\left( 
\begin{matrix}
1&1&I \cr m_d'&-m_d&M \cr
\end{matrix} \right)\nonumber\\
&& 
\sum_{s m_s}t^*_{s m_s 1 m_d'}(q,\, \theta_q,\, \theta_p,\, \phi_{pq})
\,t_{s m_s -1 m_d}(q,\, \theta_q,\, \theta_p,\,
\phi_{pq})\,,\\
V_{IM}^\pm(q,\, \theta_q)&=&V_{IM}^1(q,\, \theta_q)\pm
V_{IM}^{-1}(q,\, \theta_q)\,,
\eeqa
with 
\beqa
V_{IM}^\mu(q,\, \theta_q)&=&
\int d\,\Omega_{p}\,v_{IM}^\mu(q,\, \theta_q,\, \theta_p,\,
\phi_{pq})\nonumber\\&=& 
\frac{\hat I}{\sqrt{3}}\,\int d\,\Omega_{p}\,c(\omega,q,\Omega_q,\Omega_p)\,
\sum_{m_d m_d'}(-)^{1-m_d}
\left( 
\begin{matrix}
1&1&I \cr m_d'&-m_d&M \cr
\end{matrix} \right)\nonumber\\
&& 
\sum_{s m_s}t^*_{s m_s \mu m_d'}(q,\, \theta_q,\, \theta_p,\, \phi_{pq})
\,t_{s m_s \mu m_d}(q,\, \theta_q,\, \theta_p,\,
\phi_{pq})\,. 
\eeqa
Using now the property (\ref{vminus}), one finds with the help of
\beq
\int_0^{2\pi} d\phi_p f(-\phi_{pq})=\int_0^{2\pi} d\phi_p f(\phi_{pq})
\eeq
for a periodic function $f(\phi_{pq}+2\pi)=f(\phi_{pq})$, the relation
\beq
V_{IM}^{-1}(q,\, \theta_q)=\int d\,\Omega_{p}\,v_{IM}^{-1}(q,\,
\theta_q,\, \theta_p,\,\phi_{pq})=(-)^I\,\int
d\,\Omega_{p}\,v_{IM}^1(q,\, \theta_q,\, 
\theta_p,\, -\phi_{pq})^*=(-)^I\,V_{IM}^1(q,\, \theta_q)^*\,,
\eeq
and thus
\beq
V_{IM}^\pm(q,\, \theta_q)=V_{IM}^1(q,\, \theta_q)\pm
(-)^I\,V_{IM}^1(q,\, \theta_q)^* \,.\label{VIM}
\eeq
Correspondingly, using (\ref{wima}) one obtains
\beq
W_{IM}(q,\, \theta_q)^*=(-)^I\,\int d\,\Omega_{p}\,w_{IM}^1(q,\,
\theta_q,\, \theta_p,\,-\phi_{pq})=(-)^I\,W_{IM}(q,\, \theta_q)\,.\label{WIM} 
\eeq
From the two foregoing equations we can conclude that 
$V_{IM}^+$ and $W_{IM}$ are real for $I=0$ and 2 and 
imaginary for $I=1$, whereas $V_{IM}^-$ is imaginary for
$I=0$ and 2 and real for $I=1$. Therefore, according to (\ref{tau0c})
through (\ref{sigmal}) the following integrated asymmetries vanish
\beqa
\int d\,\Omega_{p}\,\tau_{IM}^\alpha&=&0\,\,\mbox{ for
}\,\,
\left\{\begin{array}{ll}
\alpha\in\{0,l\},&\,\,\mbox{and}\,\,I=1\\
\alpha\in\{c\},&\,\,\mbox{and}\,\,I=0,2\\
\end{array}\right\}\,,\\
\int d\,\Omega_{p}\,\sigma_{IM}^\alpha&=&0\,\,\mbox{ for
}\,\,
\left\{\begin{array}{ll}
\alpha\in\{0,l\},&\,\,\mbox{and}\,\,I=0,2\\
\alpha\in\{c\},&\,\,\mbox{and}\,\,I=1\\
\end{array}\right\}\,.
\eeqa
Instead of using these results for deriving from~(\ref{diffcross}) the
three-fold semi-exclusive differential cross section, we prefer to
start from the expression in (\ref{diffcrossa}), and obtain 
\beqa
\frac{d^3\sigma}{ dq d\Omega_q}&=&
\sum_{I}P^d_I \,
\Big\{ \sum_{M=0}^I \frac{1}{1+\delta_{M0}}\,d^I_{M0}(\theta_d)
\Re e \,[e^{iM\phi_{qd}}\,(V_{IM}^+ +P^\gamma_c\,V_{IM}^-)]
+\,P^\gamma_l\,\sum_{M=-I}^I d^I_{M0}(\theta_d)\,
\Re e \,[e^{i\psi_M}W_{IM}]
\Big\}\,.\label{diffcross_incla}
\eeqa
This expression can be simplified using the fact that
$i^{\delta_{I1}}\,W_{IM}$, $i^{\delta_{I1}}\,V_{IM}^+$ and
$i^{1-\delta_{I1}}\,V_{IM}^-$ are real according to (\ref{VIM}) and 
(\ref{WIM}). The latter two quantities can be written as 
\beqa
i^{\delta_{I1}}\,V_{IM}^+&=&2\,\Re e\,(i^{\delta_{I1}}\,V_{IM}^1)\,,\\
i^{1-\delta_{I1}}\,V_{IM}^-&=&2\,\Re e\,(i^{1-\delta_{I1}}\,V_{IM}^1)=-2\,\Im m\,(i^{-\delta_{I1}}\,V_{IM}^1)\,.
\eeqa
Using now
\beqa
\Re e \,[e^{iM\phi_{qd}}\,V_{IM}^+]&=&
\Re e \,[e^{i(M\phi_{qd}-\delta_{I1}\,\pi/2)}\,i^{\delta_{I1}}\,V_{IM}^+]=
2\,\Re e\,(i^{\delta_{I1}}\,V_{IM}^1)\,\cos[M\phi_{qd}-\delta_{I1}\,\pi/2]\,,\\
\Re e \,[e^{iM\phi_{qd}}\,V_{IM}^-]&=&
\Re e \,[\frac{1}{i}\,e^{i(M\phi_{qd}+\delta_{I1}\,\pi/2)}\,
i^{1-\delta_{I1}}\,V_{IM}^-]=
-2\,\Im m\,(i^{-\delta_{I1}}\,V_{IM}^1)\,
\sin[M\phi_{qd}+\delta_{I1}\,\pi/2]\,,\\
\Re e \,[e^{i\psi_M}\,W_{IM}]&=&
\Re e \,[e^{i(\psi_M-\delta_{I1}\,\pi/2)}\,i^{\delta_{I1}}\,W_{IM}]=
i^{\delta_{I1}}\,W_{IM}\,\cos[\psi_M-\delta_{I1}\,\pi/2]\,,
\eeqa
we find as final form for the three-fold semi-exclusive differential 
cross section 
\beqa
\frac{d^3\sigma}{dq d\Omega_q}&=&
\frac{d^3\sigma_0}{dq d\Omega_q}
\Big[1+P^\gamma_l\,\Big\{\widetilde \Sigma^l\,\cos 2\phi_q
+\sum_{I=1}^{2} P^d_I \,\sum_{M= -I}^I 
\widetilde T_{IM}^l\cos[\psi_M-\delta_{I1}\,\pi/2]
\,d^I_{M0}(\theta_d)\Big\}\nonumber\\&&
+\sum_{I=1}^{2} P^d_I \,\sum_{M= 0}^I
\Big(\widetilde T_{IM}^0\cos[M\phi_{qd}-\delta_{I1}\,\pi/2]
+P^\gamma_c\,\widetilde T_{IM}^c\sin[M\phi_{qd}+\delta_{I1}\,\pi/2]\Big)
\,d^I_{M0}(\theta_d)\Big]\,.\label{diffcrossc}
\eeqa
Here the unpolarized cross section and the asymmetries are given by
\beqa
\frac{d^3\sigma_0}{dq d\Omega_q}&=&
V_{00}^1(q,\, \theta_q)\,,\label{unpoldiff}\\
\widetilde \Sigma^l(q,\, \theta_q)\,\frac{d^3\sigma_0}{dq d\Omega_q}&=&
W_{00}(q,\, \theta_q)\,,\label{sigasy}\\
\widetilde T_{IM}^0(q,\, \theta_q)\,\frac{d^3\sigma_0}{dq d\Omega_q}&=&
(2-\delta_{M0})\, \Re e\, [i^{\delta_{I1}}\,
V_{IM}^1(q,\, \theta_q)]\,,\quad\mbox{for }0\leq M\leq I\,,
\label{tim}\\
\widetilde T_{IM}^c(q,\, \theta_q)\,\frac{d^3\sigma_0}{dq d\Omega_q}&=&
-(2-\delta_{M0})\, \Im m\, [i^{-\delta_{I1}}\,
V_{IM}^1(q,\, \theta_q)]\,,\quad\mbox{for }0\leq M\leq I\,,
\label{timc}\\
\widetilde T_{IM}^l(q,\, \theta_q)\,\frac{d^3\sigma_0}{dq d\Omega_q}&=&
i^{\delta_{I1}}\,W_{IM}(q,\, \theta_q)\,,
\quad\mbox{for }-I\leq M\leq
I\,.\label{timl} 
\eeqa
Because $V^1_{I0}$ is real according to (\ref{vw}),
the asymmetries $\widetilde T_{10}$ and $\widetilde T_{20}^c$ vanish
identically. 
We would like to point out that in forward and backward pion
emission, i.e.\ for $\theta_q=0$ and $\pi$, the following asymmetries
have to vanish
\beq
\widetilde \Sigma^l=0\,,\quad \widetilde T_{IM}^{0,c}=0\,\,\mbox{for}\,\,
M\neq 0,\quad\mbox{and}\quad  T_{IM}^{l}=0\,\,\mbox{for}\,\,M\neq 2\,,
\label{asym0}
\eeq
because in that case the differential cross section cannot depend on
$\phi_q$, since at $\theta_q=0$ or $\pi$ the azimuthal angle $\phi_q$ is
undefined or arbitrary. This feature can also be shown by straightforward
evaluation of $V_{IM}^\mu$ and $W_{IM}$ using
the explicit representation of the $t$-matrix in (\ref{smallt}). One
finds 
\beq
V_{IM}^\mu(q,\, \theta_q=0/\pi,\, \theta_p,\, \phi_{pq})=0
\quad \mbox{for }\,M\neq 0\quad\mbox{and}
\quad W_{IM}(q,\, \theta_q=0/\pi,\, \theta_p,\, \phi_{pq})=0
\quad \mbox{for }\,M\neq 2\,.\label{theta0}
\eeq 
The authors of~\cite{DaS05} were not aware of this general kinematic 
property because they evaluate the asymmetries numerically for 
$\theta_q=0$ and $\pi$ and find that the obtained values are of the order 
of $10^{-3}$. They conclude in the case of $T_{11}$ that it vanishes there but
point out that $\Sigma^l$ does not vanish. 
For completeness and also in view of the numerous errors
in~\cite{Dar04a,Dar05a,Dar05b,DaS05}, we list in the appendix~\ref{appa} the
explicit expressions of the asymmetries in terms of the $t$-matrix
elements. 

In case that only the direction of the outgoing pion is measured 
and not its momentum, the corresponding differential cross section
$d^2\sigma/d\Omega_q$ is given by an expression formally analogous to 
(\ref{diffcrossc}) where only the above asymmetries are integrated
over the pion momentum, i.e., by the replacements
\beqa
\frac{d^3\sigma_0}{dq d\Omega_q}\,&\rightarrow&\frac{d^2\sigma_0}{d\Omega_q}=
\int_{q_{min}(\theta_q)}^{q_{max}(\theta_q)} dq\,
\frac{d^3\sigma_0}{dq d\Omega_q}\,,\\
\frac{d^3\sigma_0}{dq d\Omega_q}\,\widetilde \Sigma^l(q,\, \theta_q)
&\rightarrow&\frac{d^2\sigma_0}{d\Omega_q}\,\Sigma^l(\theta_q)=
\int_{q_{min}(\theta_q)}^{q_{max}(\theta_q)} dq\,
\frac{d^3\sigma_0}{dq d\Omega_q}\,\widetilde \Sigma^l(q,\, \theta_q)\,,\\
\frac{d^3\sigma_0}{dq d\Omega_q}\,\widetilde T_{IM}^\alpha(q,\, \theta_q)
&\rightarrow&\frac{d^2\sigma_0}{d\Omega_q}\,T_{IM}^\alpha(\theta_q)=
\int_{q_{min}(\theta_q)}^{q_{max}(\theta_q)} dq\,
\frac{d^3\sigma_0}{dq d\Omega_q}\,
\widetilde T_{IM}^\alpha(q,\, \theta_q)\,,\quad \alpha\in\{0,l,c\}\,.
\eeqa
The upper and lower integration limits are given by
\beqa
q_{max}(\theta_q)&=& \frac{1}{2b}\,\Big(a\,\omega \cos \theta_q
+ E_{\gamma d}\sqrt{a^2-4b\,m_\pi^2}\Big)\,,\\
q_{min}(\theta_q)&=& \max\{0,\frac{1}{2b}\,\Big(a\,\omega \cos \theta_q
- E_{\gamma d}\sqrt{a^2-4b\,m_\pi^2}\Big)\}\,,
\eeqa
where
\beqa
a&=&W_{\gamma d}^2+m_\pi^2-4\,m_N^2\,,\\
b&=&W_{\gamma d}^2+\omega^2\sin^2\theta_q\,,\\
W_{\gamma d}^2&=&m_d(m_d+2\,\omega )\,,\\
E_{\gamma d}&=&m_d+\omega \,.
\eeqa

The general total cross section is obtained from~(\ref{diffcrossc})
by integrating over $q$ and $\Omega_q$ resulting in
\beq
\sigma(P^\gamma_l,P^\gamma_c,P^d_1,P^d_2)
= \sigma_0\,\Big[1+P^d_2\,\overline T_{20}^{\,0}\,\frac{1}{2}
(3\cos^2\theta_d-1)
+P^\gamma_c\,P^d_1\,\overline T_{10}^{\,c}\,\cos\theta_d
+P^\gamma_l\,P^d_2 \,\overline T_{22}^{\,l}\cos(2\phi_d)
\,\frac{\sqrt{6}}{4}\sin^2\theta_d\Big]\,,
\eeq
where the unpolarized total cross section and the corresponding asymmetries
are given by
\beqa
\sigma_0&=&\int d\Omega_q \int_{q_{min}(\theta_q)}^{q_{max}(\theta_q)}
dq\,\frac{d^3\sigma_0}{dq d\Omega_q}\,,\\
\sigma_0\,\overline T_{IM}^{\,\alpha}&=&
\int d\Omega_q \int_{q_{min}(\theta_q)}^{q_{max}(\theta_q)}
dq\,\frac{d^3\sigma_0}{dq d\Omega_q}\,\widetilde T_{IM}^{\,\alpha}\,,
\eeqa
with $\alpha\in\{0,l,c\}$.

Finally, we would like to point out that for coherent photoproduction
of $\pi^0$ on the deuteron formally the same expression as in
(\ref{diffcrossc}) holds with unpolarized differential cross section
and asymmetries $\Sigma^l(\theta_q)$, $T_{IM}(\theta_q)$, and
$T_{IM}^{c/l}(\theta_q)$, which are defined in analogy to
(\ref{unpoldiff}) through (\ref{timl}) with the replacements
\beqa
V_{IM}^1&\rightarrow& c(\omega,\Omega_q)\,
\frac{\hat I}{\sqrt{3}}\,\sum_{m_d m_d'}(-)^{1-m_d}
\left( 
\begin{matrix}
1&1&I \cr m_d'&-m_d&M \cr
\end{matrix} \right)
\sum_{m_d''}t^*_{m_d'' 1 m_d'}(\theta_q)
\,t_{m_d'' 1 m_d}(\theta_q)\,,\\
W_{IM}&\rightarrow& -c(\omega,\Omega_q)\,
\frac{\hat I}{\sqrt{3}}\,\sum_{m_d m_d'}(-)^{1-m_d}
\left( 
\begin{matrix}
1&1&I \cr m_d'&-m_d&M \cr
\end{matrix} \right)
\sum_{m_d''}t^*_{m_d'' 1 m_d'}(\theta_q)
\,t_{m_d'' -1 m_d}(\theta_q)\,.
\eeqa
Here, $c(\omega,\Omega_q)$ denotes a kinematic factor. A complete listing of
all polarization observables including recoil polarization of the final 
deuteron can be found in~\cite{BlB94}.

\section{Conclusions}
In this work we have derived formal expressions for the differential
cross section of incoherent pion photoproduction on the deuteron 
including various polarization asymmetries with respect to polarized 
photons and deuterons. Obviously, these expressions are generally 
valid for pseudoscalar meson production. We did not consider polarization
analysis of the final state, i.e.\ spin analysis of one or 
both outgoing nucleons. In this
case one has to evaluate instead of (\ref{diffcrossx})
\beqa
P_\alpha(j)\frac{d^5\sigma}{dq d\Omega_q d\Omega_p}&=&
c(\omega,q,\Omega_q,\Omega_p)\,
tr(T^\dagger\sigma_\alpha(j) T\rho_i)\,,\label{polj}
\eeqa
for the polarization of the ``jth'' outgoing nucleon, or
\beqa
P_{\alpha_1\alpha_2}\frac{d^5\sigma}{dq d\Omega_q d\Omega_p}&=&
c(\omega,q,\Omega_q,\Omega_p)\,
tr(T^\dagger\sigma_{\alpha_1}(1)\sigma_{\alpha_2}(2) T\rho_i)\,,\label{polij}
\eeqa
for the polarization of both outgoing nucleons. For the evaluation 
of these expressions one can proceed straightforwardly as has been 
done in~\cite{Are88}. In a subsequent paper~\cite{FiA05}, 
we will investigate the
influence of $NN$- and $\pi N$-rescattering on the various asymmetries 
of the semi-exclusive differential cross section of incoherent pion
photoproduction on the deuteron.

\acknowledgments                                                      
We would like to thank Michael Schwamb for interesting discussions and a
careful reading of the manuscript.
This work was supported by the Deutsche Forschungsgemeinschaft (SFB 443).

\appendix*
\renewcommand{\theequation}{A\arabic{equation}}
\setcounter{equation}{0}
\section{Explicit expressions for the various polarization asymmetries}
\label{appa}
We list here the explicit hermitean, bilinear forms in terms of the 
$t$-matrix elements for cross section and the various asymmetries:
\begin{enumerate}
\item The semi-exclusive differential cross section
\beq
\frac{d^3\sigma_0}{dq d\Omega_q}=\frac{1}{3}\,\int d\,\Omega_{p}\,
c(\omega,q,\Omega_q,\Omega_p)\,\sum_{s m_s m_d}|t_{s m_s 1m_d}|^2\,.
\eeq
\item The photon asymmetry for linearly polarized photons and
unpolarized deuterons
\beq
\widetilde \Sigma^l\,\frac{d^3\sigma_0}{dq
d\Omega_q}=-\frac{1}{3}\,\int d\,\Omega_{p}\, 
c(\omega,q,\Omega_q,\Omega_p)\,\sum_{s m_s m_d}
t_{s m_s 1 m_d}^*\,t_{s m_s -1 m_d}\,.
\eeq
\item The target asymmetry for vector polarized deuterons and
unpolarized photons
\beq
\widetilde T_{11}^0\,\frac{d^3\sigma_0}{dq d\Omega_q}=
\sqrt{\frac{2}{3}}\,\int d\,\Omega_{p}\,
c(\omega,q,\Omega_q,\Omega_p)\,\Im m \,\sum_{s m_s}
(t_{s m_s 1-1}^*\,t_{s m_s 10}+t_{s m_s 10}^*\,t_{s m_s 1 1})\,.
\eeq
\item The target asymmetries for tensor polarized deuterons and
unpolarized photons
\beqa
\widetilde T_{20}^0\,\frac{d^3\sigma_0}{dq d\Omega_q}&=&
\frac{1}{3\sqrt{2}}\,\int d\,\Omega_{p}\,
c(\omega,q,\Omega_q,\Omega_p)\,\sum_{s m_s}(|t_{s m_s 1 1}|^2
+|t_{s m_s 1 -1}|^2-2\,|t_{s m_s 1 0}|^2)\,,\\
\widetilde T_{21}^0\,\frac{d^3\sigma_0}{dq d\Omega_q}&=&
\sqrt{\frac{2}{3}}\,\int d\,\Omega_{p}\,
c(\omega,q,\Omega_q,\Omega_p)\,\Re e\,\sum_{s m_s}
\,(t_{s m_s 1 -1}^*\,t_{s m_s 1 0}-t_{s m_s 1 0}^*\,t_{s m_s 1 1})\,,\\
\widetilde T_{22}^0\,\frac{d^3\sigma_0}{dq d\Omega_q}&=&
\frac{2}{\sqrt{3}}\,\int d\,\Omega_{p}\,
c(\omega,q,\Omega_q,\Omega_p)\,\Re e\,\sum_{s m_s}
\,t_{s m_s 1 -1}^*\,t_{s m_s 1 1}\,.
\eeqa
\item The beam-target asymmetries for circularly polarized photons and
vector polarized deuterons
\beqa
\widetilde T_{10}^c\,\frac{d^3\sigma_0}{dq d\Omega_q}&=&
\frac{1}{\sqrt{6}}\,\int d\,\Omega_{p}\,
c(\omega,q,\Omega_q,\Omega_p)\,\sum_{s m_s}(|t_{s m_s 11}|^2-|t_{s m_s
1-1}|^2) \,,\\
\widetilde T_{11}^c\,\frac{d^3\sigma_0}{dq d\Omega_q}&=&
-\sqrt{\frac{2}{3}}\,\int d\,\Omega_{p}\,
c(\omega,q,\Omega_q,\Omega_p)\,\Re e\,\sum_{s m_s}
(t_{s m_s 1-1}^*\,t_{s m_s 10}+t_{s m_s 10}^*\,t_{s m_s 1 1})\,.
\eeqa
\item The beam-target asymmetries for circularly polarized photons and
tensor polarized deuterons
\beqa
\widetilde T_{21}^c\,\frac{d^3\sigma_0}{dq d\Omega_q}&=&
\sqrt{\frac{2}{3}}\,\int d\,\Omega_{p}\,
c(\omega,q,\Omega_q,\Omega_p)\,\Im m\,\sum_{s m_s}
(t_{s m_s 1 0}^*\,t_{s m_s 1 1}-t_{s m_s 1 -1}^*\,t_{s m_s 1 0})\,,\\
\widetilde T_{22}^c\,\frac{d^3\sigma_0}{dq d\Omega_q}&=&
-\frac{2}{\sqrt{3}}\,\int d\,\Omega_{p}\,
c(\omega,q,\Omega_q,\Omega_p)\,\Im m\,\sum_{s m_s}
t_{s m_s 1 -1}^*\,t_{s m_s 1 1}\,.
\eeqa
\item The beam-target asymmetries for linearly polarized photons and
vector polarized deuterons
\beqa
\widetilde T_{10}^l\,\frac{d^3\sigma_0}{dq d\Omega_q}&=&
\sqrt{\frac{2}{3}}\,\int d\,\Omega_{p}\,
c(\omega,q,\Omega_q,\Omega_p)\,\Im m\,\sum_{s m_s}
(t_{s m_s 1 1}^*\,t_{s m_s -1 1})\,,\\
\widetilde T_{11}^l\,\frac{d^3\sigma_0}{dq d\Omega_q}&=&
-\sqrt{\frac{2}{3}}\,\int d\,\Omega_{p}\,
c(\omega,q,\Omega_q,\Omega_p)\,\Im m\,\sum_{s m_s}
(t_{s m_s 1 -1}^*\,t_{s m_s -1 0})\,,\\
\widetilde T_{1-1}^l\,\frac{d^3\sigma_0}{dq d\Omega_q}&=&
\sqrt{\frac{2}{3}}\,\int d\,\Omega_{p}\,
c(\omega,q,\Omega_q,\Omega_p)\,\Im m\,\sum_{s m_s}
(t_{s m_s 1 1}^*\,t_{s m_s -1 0})\,.
\eeqa
\item The beam-target asymmetries for linearly polarized photons and
tensor polarized deuterons
\beqa
\widetilde T_{20}^l\,\frac{d^3\sigma_0}{dq d\Omega_q}&=&
\frac{\sqrt{2}}{3}\,\int d\,\Omega_{p}\,
c(\omega,q,\Omega_q,\Omega_p)\,\Re e\,\sum_{s m_s}
(t_{s m_s 1 0}^*\,t_{s m_s -1 0}-t_{s m_s 1 1}^*\,t_{s m_s -1 1})\,,\\
\widetilde T_{21}^l\,\frac{d^3\sigma_0}{dq d\Omega_q}&=&
\sqrt{\frac{2}{3}}\,\int d\,\Omega_{p}\,
c(\omega,q,\Omega_q,\Omega_p)\,\Re e\,\sum_{s m_s}
(t_{s m_s 1 0}^*\,t_{s m_s -1 1})\,,\\
\widetilde T_{2-1}^l\,\frac{d^3\sigma_0}{dq d\Omega_q}&=&
\sqrt{\frac{2}{3}}\,\int d\,\Omega_{p}\,
c(\omega,q,\Omega_q,\Omega_p)\,\Re e\,\sum_{s m_s}
(t_{s m_s 1 0}^*\,t_{s m_s -1 -1})\,,\\
\widetilde T_{22}^l\,\frac{d^3\sigma_0}{dq d\Omega_q}&=&
-\frac{1}{\sqrt{3}}\,\int d\,\Omega_{p}\,
c(\omega,q,\Omega_q,\Omega_p)\,\sum_{s m_s}
t_{s m_s 1 -1}^*\,t_{s m_s -1 1}\,,\\
\widetilde T_{2-2}^l\,\frac{d^3\sigma_0}{dq d\Omega_q}&=&
-\frac{1}{\sqrt{3}}\,\int d\,\Omega_{p}\,
c(\omega,q,\Omega_q,\Omega_p)\,\sum_{s m_s}
t_{s m_s 1 1}^*\,t_{s m_s -1 -1}\,.
\eeqa
\end{enumerate}


\begin{thebibliography}{99}

\bibitem{DaA03a}
E.M. Darwish, H. Arenh\"ovel, and M. Schwamb, Eur. Phys. J. A 
{\bf 16}, 111 (2003).

\bibitem{LEGS}
A. Sandorfi for the LEGS-Collaboration, private communication.

\bibitem{A2}
P. Pedroni for the A2-Collaboration, private communication.

\bibitem{LeP96}
M.I. Levchuk, V.A. Petrun'kin, and M. Schumacher, 
Z. Phys. A {\bf 355}, 317 (1996).   

\bibitem{DaA03b}
E.M. Darwish, H. Arenh\"ovel, and M. Schwamb, Eur. Phys. J. A 
{\bf 17}, 513 (2003).

\bibitem{ArF04}
H. Arenh\"ovel, A. Fix, and M. Schwamb, Phys.\ Rev.\ Lett. 
{\bf 93}, 202301 (2004).

\bibitem{LoS00}
A. Loginov, A.Sidorov, and V. Stibunov, 
Phys. Atom. Nucl. {\bf 63},391 (2000) (Yad. Fiz. {\bf 63}, 459 (2000)).

\bibitem{Dar04a}
E.M. Darwish, J. Phys. G {\bf 31}, 105 (2004).

\bibitem{Dar05a}
E.M. Darwish, Nucl. Phys. A {\bf 735}, 200 (2005).

\bibitem{Dar05b}
E.M. Darwish, Nucl. Phys. A {\bf 748}, 596 (2005).

\bibitem{Dar05c}
E.M. Darwish, nucl-th/0504031 

\bibitem{DaS05}
E.M. Darwish and A. Salam, nucl-th/0505002.

\bibitem{Are88}
H. Arenh\"ovel, Few-Body Syst.\ {\bf 4}, 55 (1988).  
	
\bibitem{BlB94}
F. Blaazer, B.L.G. Bakker, and H.J. Boersma, 
Nucl. Phys. A {\bf 568}, 681 (1994).

\bibitem{FiA05}
A. Fix and H. Arenh\"ovel, in preparation.

\bibitem{Ros57}
E.M. Rose, {\it Elementary Theory of Angular Momentum}, Wiley New York 1957.

\bibitem{Rob74} 
B.A. Robson, {\it The Theory of Polarization Phenomena}, 
Clarendon Press, Oxford 1974.

\end{thebibliography}
\end{document}